\documentstyle[aps,preprint]{revtex}

\begin{document}
%\twocolumn[
\tighten

%\draft

\preprint{WM-95-105}

\title {\bf Electroproduction and Hadroproduction of Light Gluinos}

\author{C. E. Carlson, G. D. Dorata, D. Morgan, and M. Sher}
\address{Physics Department, College of William and Mary,
Williamsburg, VA 23187, USA}

\maketitle

%\vskip -0.2in

\begin{abstract} %\widetext

In a class of supergravity models, the gluino and photino are massless at tree
level and
receive small masses through radiative corrections.  In such models, one
expects a
gluino-gluon bound state, the $R_0$, to have a mass of between 1.0 and 2.2 GeV
and a
lifetime between $10^{-10}$ and $10^{-6}$ seconds.  Applying perturbative QCD
methods
(whose validity we discuss), we calculate the production cross sections of
$R_0$'s in
$e-p$, $\pi-p$, $K-p$, $\overline{p}-p$ and $p-p$ collisions.  Signatures are
also
discussed.

\end{abstract}

\vskip 0.2in


%]

%\newpage

\narrowtext

\section{Introduction}

In searches for supersymmetric particles, it is generally assumed
that the masses of the new particles are $O(100-1000)$ GeV, and
thus they can only be produced in high energy accelerators.
However, a possibility which has been receiving increasing
attention of
late\cite{oldpapers,newer,gfone,banks,gftwo,gfthree,masiero,kolb,gffour}
is that the gluino and photino are extremely light, with masses in range
of hundreds of MeV. If so, then the gluino--gluon bound state, called the
glueballino (which we designate as $R_0$), would have a mass in the $1-2$
GeV range, and would be very long-lived, possibly with a lifetime as long
as that of the muon.  The possibility that a  strongly interacting,
long-lived particle with a mass only slightly greater than that of the
neutron could have evaded detection is astonishing, and yet  this appears
to be the case:  An $R_0$ mass between 1.0 and 2.2 GeV would not yet have
been experimentally excluded\cite{gfone}.

Why would one expect gluinos to be so light?  The fact that
scalar quark masses must be greater than the $W$ mass shows that
supersymmetry is broken at the scale of at least $O(100)$ GeV.
However, the source of gaugino masses in many supergravity models
is completely different from the source of scalar masses, since
the former arise from dimension-3 SUSY-breaking operators.  In
some such models, such as those in which SUSY is broken in the
hidden sector and there are no gauge singlets\cite{banks,gftwo,gfthree},
the dimension-3 SUSY-breaking terms are either absent or suppressed by a
factor of the Planck mass.  Thus, in these models, the gluino and
photino\footnote{When we say photino in this paper, we are
actually referring to the lightest neutralino.  However, in
models in which the lightest neutralino is extremely light, it
tends to be a pure photino state.} are massless at tree level.
Masses will be generated by radiative corrections; these were
calculated by Farrar and Masiero\cite{masiero}, who found that as the
typical SUSY breaking scale varies from 100 to 400 GeV, the gluino mass
decreases from 700 to 100 MeV, as the photino mass increases from
approximately 400 to 900 MeV. Although the photino might, in
these models, be somewhat heavier than the gluino, the lightest
color-singlet containing the gluino, the $R_0$, will be heavier
than the photino, for the same reason that a glueball, comprised
of massless gluons, has a mass in the $1-2$ GeV range. In fact,
if the gluino is light, then the $R_0$ mass should be very
similar to that expected for the lightest $0^{++}$ glueball, i.e.
$1.4\pm.4$ GeV.

 If this is the case, then the photino will then be stable, and an
ideal candidate for the dark matter.  In fact, Farrar and Kolb\cite{kolb}
have shown that if the ratio of the $R_0$ mass to the photino
mass is in the range from 1.2 to 2.2, then the relic abundance of
photinos is just right to account for the dark matter; this mass
range overlaps nicely with the range of masses calculated from
radiative corrections.   Since the gluino will decay through
virtual scalar quark processes, the $R_0$ lifetime should be
quite long; estimates range from
$10^{-10}$ to
$10^{-6}$ seconds.

 How could such a light, long-lived, strongly interacting
particle have escaped detection?\cite{gftwo,gffour}
  Missing energy searches (the
classic signatures of supersymmetry) require large transverse
missing energy, and gluinos would not have been detected if the
lifetime is greater than $10^{-10}$ seconds.  Beam dump
experiments which look for the subsequent interaction of the
photino would not be sensitive since the photino cross section is
significantly smaller (by a factor of $O(m_W/m_{sq})^4$).
Experiments at CUSB\cite{cusb} and ARGUS\cite{argus} look for radiative
$\Upsilon$ decays; these experiments can rule out a region of gluino masses
which correspond to $R_0$ masses
from roughly $2$ to
$4$ GeV, for any lifetime; other experiments modify the bounds
slightly.   These experiments are all discussed by
Farrar\cite{gftwo,gffour}, who provides a plot of the region of the
mass-lifetime plane excluded by each of these experiments; the region from
$1.0$ to $1.5$ GeV is not excluded for any lifetime, and the region from
$1.5$ to $2.2$ GeV is only excluded for lifetimes between
$10^{-6}$ and
$10^{-8}$ seconds.

There was some excitement recently\cite{newer} about the possibility that
the presence of light gluinos could alter the running of the QCD
coupling constant between $Q^2=m^2_{\tau}$ and $Q^2=m^2_Z$.  It
appears that the value of the QCD coupling at the smaller scale
is too high, given its value of the larger scale, and modifying
the beta function by inclusion of light gluinos could account for
the discrepancy.  However, it has been pointed out\cite{newer} that the
uncertainties in this analysis are large, and that the data, at
present, can not be used to either establish or rule out light
gluinos.  Similar arguments apply to jet production at Fermilab
and LEP; the uncertainties are too large. In addition, an
additional state at $1.4$ GeV has been seen, which could be a
gluino-gluino bound state, but distinguishing such a state from
other possible exotics, such as hybrids, will not be easy.

In order for experimenters to probe the allowed mass and lifetime
range, it is necessary to have reasonably accurate values for the
production cross section of gluinos.  This is not always easy.
For example, Farrar\cite{gffour} has proposed dearching for $R_0$ decays
into
$\eta+\tilde{\gamma}$ by looking for
$\eta$'s in high-intensity kaon beams.  This could certainly
establish the existence of gluinos, but the production of $R_0$'s
relative to kaons cannot be calculated in perturbative QCD, due
to the fact that neutral kaon beams are produced at low
transverse momentum.  On the other hand, one can compute gluino
production cross sections at high $p_T$ reliably.  The cross
section for photoproduction of gluino pairs was calculated\cite{cebaf}
recently, with the hope of using the photon tagger in the Large
Acceptance Spectrometer at CEBAF.  Although this calculation did
not directly impose a $(p_T)_{min}$ cut, such a cut would be done
by the experimeters, and the event rates were high enough that
this cut would not lower the signal too much\footnote{The
signature for $R_0$ production in that experiment assumed very
light or massless photinos, however, and the mass range expected
from the above would likely require a different signature.
Signatures of $R_0$ production will be discussed later.}. As
pointed out there, the long $R_0$ lifetime and relatively light
mass indicates that high-luminosity, lower energy accelerators
will be better suited for exploring the allowed range.

In this paper, we will calculate the electroproduction and
hadroproduction cross sections for light gluino pairs.
Our primary motivation is as follows.  Searching for gluinos in high
intensity kaon beams, as suggested by Farrar, may very well be the best
way to discover gluinos if they are there.  However, the absence of a
reliably calculable  production cross section will make it difficult for
experimenters to exclude regions of the mass-lifetime plane; only regions
of the mass-lifetime-production cross section volume can be excluded.  In
electroproduction and hadroproduction, one can reliably calculate the
cross-sections in some kinematic regions, and although such experiments
may not be the best way to find gluinos, they do offer the possibility
of reliably excluding certain regions of the mass-lifetime plane (given
the uncertainties associated with pQCD, of course, which we discuss in
the next section).

 We will begin by considering electroproduction,
discussing the validity of perturbative QCD as well, and
then turn to hadroproduction, calculating cross sections for
$\pi p$, $K p$, $\overline{p}p$ and $pp$ collisions.  We will then
discuss experimental signatures of light gluinos.

\section{Electroproduction of light gluinos}

\subsection{Cross section}

The relevant diagrams for electroproduction are shown in Fig 1., and
the square of the resulting matrix element is given in the Appendix.
In integrating over phase space, the same procedure was used as in
the photoproduction calculation. The integrations over the gluino
momenta are performed in the $\vec r=0$ reference frame, and then
re-expressed in covariant form. The subsequent integration over the
outgoing quark momentum is done in the quark-photon center of
momentum frame. We do not integrate over the outgoing electron,
instead we will express our results as a differential cross section of
the form $E_{l_2} d\sigma/d^3l_2$.

Once we obtain the subprocess cross section, we must embed the
target quark in a proton and integrate over the allowed
values of $\hat s$. We fold the cross
section with the distribution functions of the quark in a proton

\begin{eqnarray}
E_{l_2} {d\sigma \over d^3l_2} &=& \int dx\, \sum_q e_q^2 f_q(x)
E_{l_2} {d\hat\sigma(\hat s) \over d^3l_2}\cr &=& \int dx\,  E_{l_2}
{d\hat\sigma(\hat s)\over d^3l_2} F_{2p}(x)/x.
\end{eqnarray}
where $F_{2p}$ is the proton electromagnetic structure function. We
used  up-to-date CTEQ distribution functions (specifically
CTEQ1L) for all of  our calculations.

Figure 2 shows the differential cross section $E_{l_2} d\sigma/d^3l_2$ plotted
vs. the
energy of the outgoing electron. The incident electron  energy is 12 GeV
(corresponding to the maximum energy likely to be
reached at CEBAF in the near future)   and
the polar angle of the outgoing elecron is fixed  at 15$^\circ$. We have
assumed that each final state gluino will be bound  within a glueballino (a
gluon/gluino bound state) and in evaluating our  formulas, we have given
the gluino an effective mass equal to the glueballino mass. Our  results
are sensitive to this mass, and we have plotted our results for
glueballino masses of 1.0, 1.2, and 1.5 GeV. The results are not  sensitive to
the quark
mass since there are no collinear singularities for spacelike $q^2$.  Our
calculations
assigned the quark an effective mass $xm_N$ so that the threshold for the
$\gamma+q$
subprocess would be at  the same photon energy as for the overall $\gamma+p$
process.
Letting the quark be massless would make a negligible difference away from
threshold.

\subsection{Applicability of pQCD}

The energies and transverse momenta involved here  are not very large and
one may worry about the validity of calculations based on perturbative
QCD.  Already our worries should be assuaged by insensitivity to the
quark mass displayed in the photoproduction calculation, where even using
a quark mass as large as 1 GeV has only a small effect on the size of the
calculated cross section.

We may study the reliability of pQCD in more detail by considering how
off shell the internal propagators are in these calculations.  Far off
shell means the internal particles can travel only short distances, and
short distances are where pQCD is valid.  Two of the three propagators in
the two diagrams of the photoproduction version
of Fig. 1 are always far off shell.  These
are the quark propagator in the s-channel diagram and the gluon propagator,
which has
to supply the energy to produce a massive gluino or even a glueballino
pair.  The quark propagator in the u-channel diagram, however, can get
rather close to singular when the photon and outgoing quark are
collinear.

We studied the importance of this near singularity in the photoproduction
case.  First, we control the singularity as we normally do by inserting a
quark mass.  Then we add an extra requirement, that $|\hat{u}|$ be
greater than some fixed amount to ensure that whatever contributions we
keep in our calculation are perturbatively reliable.  Here, the ``hat'' denotes
a
Mandelstam variable for the $\gamma-q$ subprocess. Requiring
$|\hat{u}| > 1$ GeV${}^2$ (which, if we include the quark mass, means the
propagator
is off shell by more than $1$ GeV${}^2$) leads to a decrease in cross
section of less than five percent for incoming photon energies of 10 GeV
and glueballino masses in the $1.0-1.5$ GeV range.  We conclude that the
bulk of our cross section comes from kinematics where all internal
propagators are far off shell and hence that the perturbative
calculations are good approximations to the correct cross section.

\subsection{Event rates}

The Hall B Large Acceptance Spectrometer at CEBAF can accept a luminosity
of $10^{34}$ cm${}^{-2}$ sec${}^{-1}$.  (The luminosity for Hall B is set
by what the detector can accept rather than by what the accelerator can
produce.)  For electroproduction, taking $10^{-3}$ nb/GeV${}^2$ as a
typical cross section in Fig. 2, this translates into a typical event rate
of 1 per 100 seconds.  Similar event rates will be obtained for the proposed
ELFE
accelerator, if it has a large acceptance detector.  Even with a lifetime near
the
upper end of the expected range, one microsecond, one would have an $R_0$
decaying in the
detector several times per day.  Signatures of these decays will be discussed
below.

\section{Hadroproduction of light gluinos}

We will consider two classes of hadroproduction reactions. The first class
involves reactions in which the incident particle contains one or more
valence $\overline u$ or $\overline d$ anti-quarks, including $\pi p$, $K
p$ and $\overline{p}p$. Then we will consider production via $pp$
collisions at the end of the section.

If one of the hadrons contains valence anti-quarks, then the dominant mode
of gluino production will be through $q\overline q$ annihilation (see
Fig. 3). The calculation for this process is straightforward, and the
resulting cross section is given by\cite{chls}

\begin{equation}
\hat \sigma = {16\pi\alpha_s^2\over 9 \hat s} \big({1+2M^2\over\hat s}\big)
\sqrt{1-4M^2\over \hat s}
\end{equation}
where $\hat s$ is the total energy in the quark-photon center of momentum
frame, and once again we will consider $M$ to be the glueballino mass. In
order to obtain the total cross section, we fold this subprocess cross
section with the hadron distribution functions

\begin{equation}
\sigma =\int\int \hat \sigma \sum_q \big(q_a(x_1)\overline
q_b(x_2)+\overline q_a(x_1)q_b(x_2)\big) dx_1\ dx_2\
\end{equation}
where $x_1$ and $x_2$ are the momentum fractions of the quark and
anti-quark, and $q_a(x)$ and $q_b(x)$ are the quark (anti-quark)
distribution functions for each hadron. For the proton, we once again use
the CTEQ1L distribution functions. For mesons we will use

\begin{eqnarray}
v(x)=.75 x^{-1/2} (1-x) \nonumber \\
s(x)=.12 x^{-1} (1-x)^5
\end{eqnarray}
for the valence and sea quark (or anti-quark) distribution functions,
respectively. We will also assume ``$SU(2 {1\over 2})$" for the strange sea
quarks, that is, we assume that there are half as many
$s\overline s$ pairs in the quark sea of the meson as there are $u\overline
u$ and $d\overline d$ pairs.

The results for $K^-p$, $\pi^- p$,  and $\overline{p}p$ are
shown in figures 4, 5, and 6 respectively. The total cross section
is plotted versus the incident beam energy for glueballino masses of 1.0,
1.5, and 2.0 GeV.

The cross sections are quite high.  For example, for the 18 GeV $\pi^-$
beam at Brookhaven, one has an event rate of roughly 0.5/microbarn/sec.
For a 1.0 GeV $R_0$, this gives an event every two seconds (for a microsecond
lifetime, an $R_0$ will decay within a meter of the interaction region every
hour
or so).  For a 2.0 GeV $R_0$, the rate is two orders of magnitude smaller.

The second class of hadroproduction reactions involves cases in which the
incident particle does not contain any valence anti-quarks, for example,
proton-proton collisions. Although the process $q\overline q\rightarrow
\tilde{g}\tilde{g}$ will still contribute (due to the presence of sea
antiquarks in both particles) it will be suppressed relative to the
cases in which there are valence anti-quarks.  To the same order, there will be
a contribution from gluon
fusion\cite{chls},
$gg\rightarrow
\tilde{g}\tilde{g}$. The process $gq\rightarrow
\tilde{g}\tilde{g}q$, although higher order in the coupling constant, may be
competitive with this process
 Here, the calculation is necessarily imprecise, since the gluon
distribution function we use will be modified by the presence of gluinos in the
sea (we
are omitting contributions from primordial gluinos.  Using a conventional gluon
distribution function  we have found that the two contributions are similar,
and that the
resulting curves are the same shape as those in Fig. 6, but are roughly two
orders of
magnitude smaller.

\section{SIGNATURES}

One general signature of the glueballino is that it can aspects of both a
long lived and a short lived particle.  It should, like a long lived
particle, decay a long distance away from where it was produced.  Then if
it decays into two or more hadrons, it should have a wide decay width in
the sense that the spread of mass visible in the decay should be large.
This is a consequence of the varying energy taken away by the almost
non-interacting photino if the final state is three or more particles in
total.  It is important in this case that the decay not be one that could
be mimicked by known weakly decaying particles.  With this in mind, it
was proposed to look for the decay of the glueballino into four charged
pions  plus an unobserved photino; the appearance of four charged pions
emerging from a vertex away from the interaction point would be a
``gold-plated" signature for glueballinos, and the branching ratio of known
mesons in this mass range into four charged pions is not unusually small.

This would be the best signature if the photino was very light.  However,
the work of Farrar and Masiero, and the cosmological arguments of Farrar
and Kolb, suggest that the photino is not particularly light, and thus
the decay into four charged pions will, if even kinematically allowed, be
suppressed significantly.  One could look for three pions and a photino,
with the three pions having more invariant mass than the kaon.

The two most interesting two-body decays are into $\pi^o+\tilde{\gamma}$
and, if kinematically allowed, $\eta+\tilde{\gamma}$.  It is the latter
decay that Farrar\cite{gffour} has proposed looking for in experiments that
produce kaon beams since there may be some admixture of $R_0$ in the beam;
the
$\eta$ will subsequently decay into three pions with more invariant
mass than the kaon.  Due to SU(3) factors, the branching ratio of
$\eta+\tilde{\gamma}$ will be, to the extent that the $\eta$ mass does
not suppress the rate, 10\% of the $\pi^o+\tilde{\gamma}$ ratio.
The appearance of a single $\pi^0$ a distance from the vertex may be
difficult to pick out of the background, and the $\eta$ may thus be
easier to find.

One could also look for $\pi^+\pi^-\tilde{\gamma}$ where
the pions have an invariant mass greater than the kaon. Although phase
space arguments indicate a branching ratio of $O(10)^{-3}$ \cite{gffour},
such arguments generally underestimate the multi-hadron decay rates of
mesons in the 1--2 GeV mass range; for many such mesons the multi-hadron
decay will dominate the two-body decay.  Thus, the branching ratio for
this mode could be sizable.

\section{CONCLUSIONS}

It is remarkable that the existence of a long-lived, strongly interacting
particle with a
mass just slightly above that of the neutron cannot be experimentally excluded.
Given
that such a particle is a consequence of a class of supergravity models, a
comprehensive
search for light gluinos is well-motivated.  Although the best method of
detecting
gluinos might well be to look for their presence in kaon beams, the absence of
a reliable
production cross section precludes the possibility of definitely ruling out
gluinos
in a given mass and lifetime region.  In this article, we have calculated the
rate for
electroproduction and hadroproduction of light gluinos, in a kinematic regime
in which
perturbative QCD should be fairly reliable.  The event rates are quite high,
and the
signatures fairly distinctive.  Failure to find gluinos at the predicted rate
(or within
a factor of a few, given the uncertainties in perturbative QCD at this scale)
will
definitively rule out light gluinos in a given mass-lifetime region.  Their
discovery
will revolutionize particle physics, and lead to a new generation of ``gluino
factories".

\section{ACKNOWLEDGMENT}

We thank Glennys Farrar, John Kane, Larry Weinstein and Suh-Urk Chung for many
useful
discussions.  This work was supported by the NSF
under Grant No. NSF-PHY-9306141.

\bibliographystyle{unsrt}

\section{APPENDIX}

The diagrams for electroproduction of light gluinos are shown in Fig. 1.  In
terms of the momenta defined in the diagrams, we define, with $m$ being the
mass of the
gluino,
\begin{eqnarray}
\Delta&=&(k_1-k_2)/2\nonumber\\
\Delta^2&=&m^2-r^2/4\nonumber\\
s&=&(p_1+l_1)^2\nonumber\\
t&=&(p_1-l_2)^2\nonumber\\
s_h&=&(p_2+r)^2\nonumber\\
t_h&=&(p_1-p_2)^2\nonumber\\
u_h&=&(p_1-r)^2\nonumber\\
Q&=&(l_1+l_2)/2
\end{eqnarray}
Then, the square of the matrix element is

\begin{eqnarray}
\overline{|{\cal M}|^2}
&=&
-{4e^2e_q^2g_s^4\over q^4r^4}\Biggl\{\nonumber\\ &\ &{32\over
u_h^2}\bigl[(1+{u_h\over
s_h})q^2(s_h-r^2)+(s-t-q^2-r^2+t_h-4l_2\cdot
r)^2+q^2t_h\bigr](p_1\cdot
\Delta)^2\nonumber\\
&+&{32\over
s_h^2}\bigl[(s-t)^2+q^2t_h-q^2(r^2-u_h)(1+{s_h\over
u_h})\bigr](p_2\cdot\Delta)^2\nonumber\\
&+&{64\over s_hu_h
}\bigl[q^2(s_h+t_h+u_h-2r^2)+(s-t)(s-t-q^2-r^2+t_h-4l_2\cdot r)
\bigr]p_1\cdot
\Delta p_2\cdot \Delta\nonumber\\
&-&{128\over u_h}(s-t-q^2-r^2+t_h-4l_2\cdot r)p_1\cdot\Delta
Q\cdot\Delta\nonumber\\ &-&{128\over s_h}(s-t)p_2\cdot\Delta
Q\cdot\Delta\nonumber\\
&+&{128\over
s_hu_h}\bigl[r^4-r^2(s_h+t_h+u_h)+s_hu_h\bigr](Q\cdot\Delta)^2\nonumber\\
&-&{4\over
s_h^2}(2\Delta^2+r^2)\bigl[q^2s_hu_h-r^2q^2(s_h+t_h+u_h)+r^4q^2\nonumber\\
&+&s_h(s-t)(q^2+r^2-t_h+4l_2\cdot r)-r^2(s-t)^2\bigr]\nonumber\\
&-&{4\over
u_h^2}(2\Delta^2+r^2)\bigl[q^2s_hu_h-r^2q^2(s_h+t_h+u_h)+r^4q^2\nonumber\\
&-&u_h(s-t-q^2-r^2+t_h-4l_2\cdot r)(q^2+r^2-t_h+4l_2\cdot
r)\nonumber\\ &-&r^2(s-t-q^2-r^2+t_h-4l_2\cdot
r)^2\bigr]\nonumber\\
&-&{4\over s_hu_h}\Bigl[2\Delta^2t_h(q^2+r^2-t_h+4l_2\cdot
r)^2\nonumber\\ &+&(s-t-q^2-r^2+t_h-4l_2\cdot
r)\bigl((q^2+r^2-t_h+4l_2\cdot r)[2\Delta^2(r^2-u_h)\nonumber\\
&-&r^2t_h]-r^2(s-t)(4\Delta^2-2t_h+r^2-u_h)\bigr)\nonumber\\
&+&(q^2+r^2-t_h+4l_2\cdot
r)(s-t)[2\Delta^2(s_h-r^2)+r^2t_h]\nonumber\\
\nonumber\\ &+&2r^2q^2t_h(s_h+t_h+u_h)
-r^2(r^2-u_h)(s-t-q^2-r^2+t_h-4l_2\cdot r)^2\nonumber\\
&+&(s_h-r^2)[4\Delta^2q^2(r^2-u_h)+r^2(s-t)^2]\nonumber\\
\nonumber\\ &+&2r^2(6\Delta^2+r^2)q^2t_h\Bigr]\Biggr\}
\end{eqnarray}

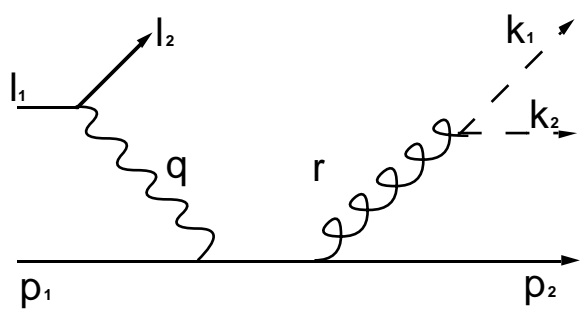
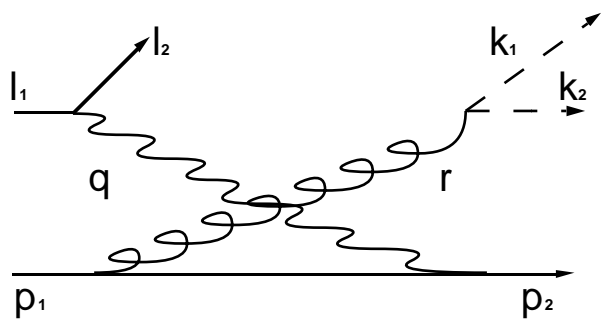
\begin{figure}

%\vglue 2.0in
%\hskip -0.2in {\special{picture feynman scaled 400}} \hfil

\caption{The Feynman diagrams for electroproduction of gluinos. The dashed
lines represent
the gluinos.}
 \label{fig1}
\end{figure}

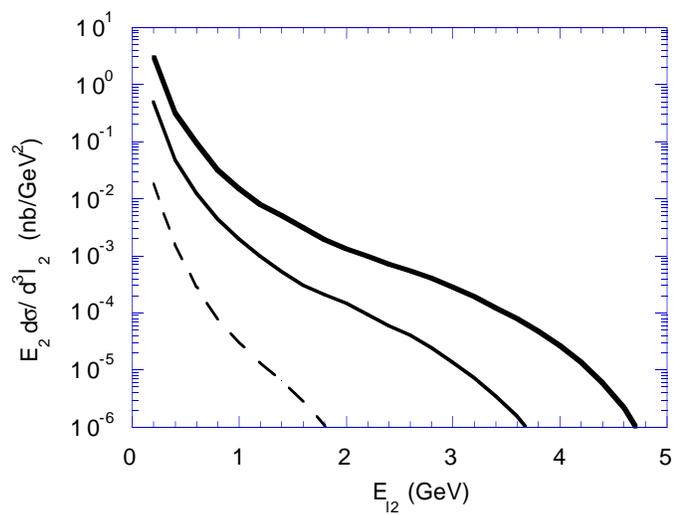
\begin{figure}

%\vglue 2.5in
%\hskip -0.2in {\special{picture electro12 scaled 700}} \hfil

\caption{The differential cross section for electroproduction of
glueballino pairs is plotted vs. the energy of the outgoing electron. The
incident electron  energy is 12 GeV and the polar angle of the outgoing
elecron is fixed  at 15$^\circ$. The solid, dashed, and dotted lines show
the results for a glueballino mass of 1.0 GeV, 1.2 GeV, and 1.5 GeV
respectively.}
\label{fig2}
\end{figure}

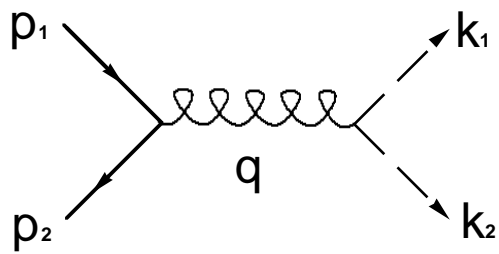
\begin{figure}

%\vglue 1.0in
%\hskip -0.2in {\special{picture feynman2 scaled 500}} \hfil

\caption{The Feynman diagram for the production of gluinos via $q\overline q$
annihilation.
The dashed lines represent the gluinos.}
\label{fig3}
\end{figure}

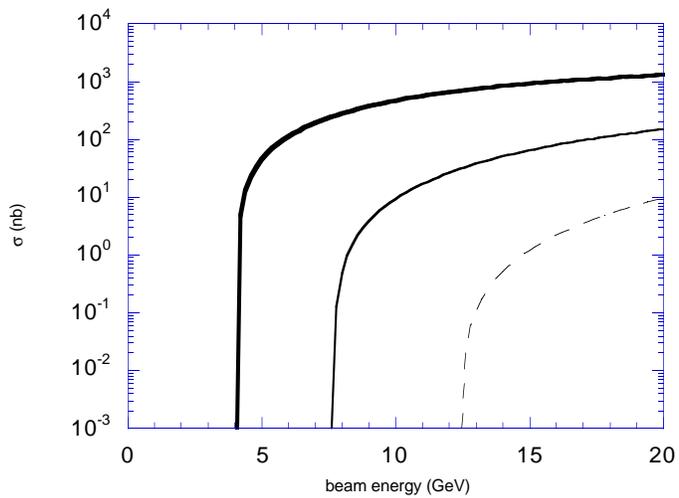
\begin{figure}
%\vglue 2.5in
%\hskip -0.2in {\special{picture kaon scaled 700}} \hfil

\caption{The total cross section for $K^-p \rightarrow
\tilde{g}\tilde{g} +X$ is plotted vs. the energy of the incident kaon beam
. We are assuming that all gluinos end up in $R_0$'s, as expected. The heavy,
thin, and
dashed lines show the results for a glueballino mass of 1.0 GeV, 1.5 GeV, and
2.0 GeV
respectively.}
\label{fig4}
\end{figure}

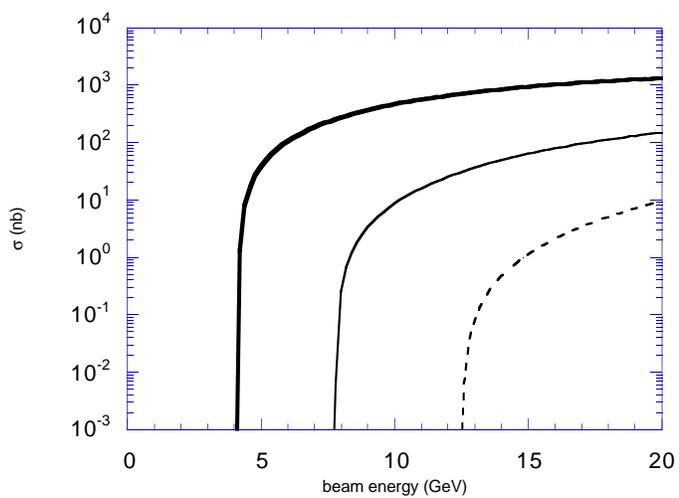
\begin{figure}
%\vglue 2.5in
%\hskip -0.2in {\special{picture pion scaled 700}} \hfil

\caption{The total cross section for $\pi^-p \rightarrow
\tilde{g}\tilde{g} +X$ is plotted vs. the energy of the incident pion beam
. The heavy, thin, and dashed lines show the results for a glueballino
mass of 1.0 GeV, 1.5 GeV, and 2.0 GeV respectively.}
\label{fig5}
\end{figure}

\begin{figure}
%\vglue 2.5in
%\hskip -0.2in {\special{picture ppbar scaled 700}} \hfil

\caption{The total cross section for $p\overline p \rightarrow
\tilde{g}\tilde{g} +X$ is plotted vs. the energy of the incident anti-proton
beam. The
heavy, thin, and dashed lines show the results for a glueballino mass of 1.0
GeV, 1.5 GeV,
and 2.0 GeV respectively.}
\label{fig6}
\end{figure}

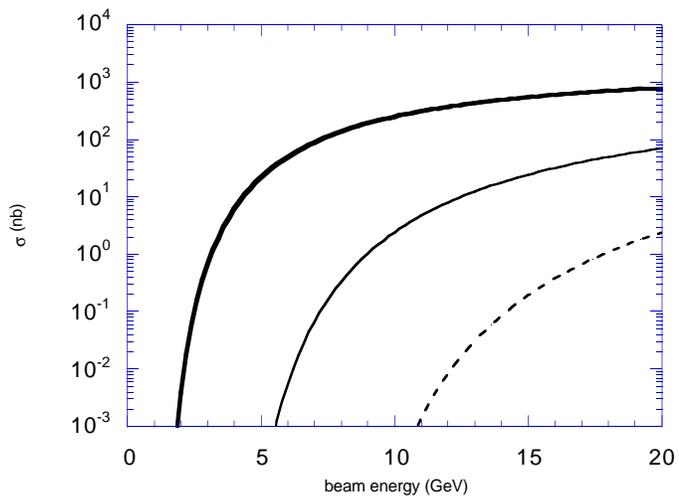
\end{document}